\pdfoutput=1
\documentclass[11pt]{article}
\usepackage{jheppub}

\usepackage{amsmath}
\usepackage{amssymb}
\usepackage{graphicx,color}
\usepackage{amsthm}
\usepackage{subfigure}

\usepackage{longtable}

\usepackage{ifpdf}

\usepackage[boxsize=2em,aligntableaux=center]{ytableau}

\usepackage{multirow}

\newcommand{\be}{\begin{equation}}
\newcommand{\ee}{\end{equation}}
\newcommand{\ba}{\begin{eqnarray}}
\newcommand{\ea}{\end{eqnarray}}
\newcommand{\nn}{\nonumber}
\newcommand{\lp}{\left(}
\newcommand{\rp}{\right)}
\newcommand{\ls}{\left[}
\newcommand{\rs}{\right]}

\renewcommand{\a}{\alpha}

\newcommand{\N}{\mathcal{N}}
\newcommand{\bZ}{\mathbb{Z}}
\newcommand{\cZ}{\mathcal{Z}}

\notoc

\title{\centering Mathieu moonshine in four dimensional $\mathcal{N}=1$ theories}

\author[a]{Timm Wrase}

\affiliation[a]{Stanford Institute for Theoretical Physics\\ Stanford University, Stanford, CA 94305, USA}

\emailAdd{timm.wrase@stanford.edu}

\abstract{We show that the recently discovered Mathieu moonshine plays a role for certain four dimensional theories with $\mathcal{N}=1$ supersymmetry. These theories are obtained from the $E_8 \times E_8$ heterotic string theory by compactifying on toroidal orbifolds. We find that a universal contribution to the holomorphic gauge kinetic function can be expanded in such a way that the expansion coefficients are the dimensions of representations of the Mathieu group M$_{24}$.
}

\begin{document}

\makeatletter
\let\old@fpheader\@fpheader
\renewcommand{\@fpheader}{\old@fpheader\hfill
SU/ITP-14/04}
\makeatother

\maketitle

\newpage

\section{Introduction}
Recently Eguchi, Ooguri and Tachikawa \cite{EOT} observed that the elliptic genus of the $K3$ manifold exhibits `Mathieu moonshine': when expanding the elliptic genus in terms of Virasoro characters of the $\mathcal{N} = 4$ superconformal algebra, they find that the first few expansion coefficients are related to the sum of dimensions of irreducible representations (irreps) of the Mathieu group M$_{24}$. This observation was further checked and confirmed in \cite{Miranda, Gaberdiel:2010ch, Gaberdiel:2010ca, Eguchi:2010fg}, before Gannon proved that all the expansion coefficients are positive sums of dimensions of irreducible M$_{24}$ representations \cite{Gannonproof}.

This Mathieu moonshine is very interesting since it points towards a deep connection between the sporadic group M$_{24}$ and the $K3$ manifold that yet needs to be understood. Since $K3$ manifolds play a crucial role in string compactifications one may ask whether this observation is also relevant for the spacetime theories one obtains from string compactifications involving $K3$. One of these connection was discussed in \cite{Hohenegger:2011us}, where the authors showed that certain BPS saturated 1-loop amplitudes in type II string theory compactified on $T^2 \times K3$ are related to the elliptic genus of $K3$ and therefore to M$_{24}$. Another connection was made in \cite{Harvey:2013mda}, where the authors studied compactifications of type II string theory on $K3 \times S^1$ in the presence of NS5-branes. There the authors find that certain BPS states are counted by a mock modular form that is closely related to Mathieu moonshine.

For the heterotic $E_8 \times E_8$ string theory compactified on $K3 \times T^2$ this question was addressed in \cite{Cheng:2013kpa}. The authors find that the four dimensional $\mathcal{N}=2$ theories have gauge theories whose couplings receive corrections that are related to the elliptic genus of $K3$ \cite{Harvey-Moore} and can always be expanded in such a way that the expansion coefficients are exactly the same as the ones appearing in Mathieu moonshine. Therefore Mathieu moonshine is clearly important for certain four dimensional $\N=2$ spacetime theories. Furthermore, it was shown in \cite{Cheng:2013kpa} that these dimensions of irreps of M$_{24}$ are related to Gromov-Witten invariants in the dual type II string theory compactified on particular $CY_3$ manifolds that are elliptic fibrations over a Hirzebruch surface. This means that Mathieu moonshine actually teaches us something about the geometry of these particular $CY_3$ manifolds.

It is thus clear that the observations of Eguchi, Ooguri and Tachikawa \cite{EOT} is not only relevant to $K3$ manifolds and certain $CY_3$ manifolds but also plays a role in four dimensional spacetime theories that preserve $\mathcal{N}=2$ supersymmetry. In this work we show that there is a large class of related $\mathcal{N}=1$ four dimensional theories whose 1-loop corrections to the gauge kinetic function can likewise be expanded in such a way that the expansion coefficients are the same dimensions of M$_{24}$ representations. We do that by recalling from \cite{Dixon1991649} that the moduli dependent corrections to the gauge kinetic function in toroidal orbifold compactifications arise only from $\mathcal{N}=2$ subsectors. There is a large class of toroidal orbifolds $T^6/\mathbb{Z}_N$ and $T^6/\mathbb{Z}_N \times \mathbb{Z}_M$ that have $\mathcal{N}=2$ subsectors which give corrections to the gauge kinetic function that are closely related to the ones found in \cite{Cheng:2013kpa}. This class of toroidal orbifolds therefore leads to four dimensional theories that preserve $\mathcal{N}=1$ supersymmetry, have a variety of different gauge groups and matter content and exhibit Mathieu moonshine in the gauge kinetic coupling. In particular, these theories include GUT-like models with $E_6$ gauge group and chiral matter in the $\underline{27}$ representation.

The outline of the paper is as follows: In section \ref{sec:setup} we discuss some basic facts about the heterotic $E_8 \times E_8$ string theory compactified on toroidal orbifolds. Then we discuss in section \ref{sec:main} how the 1-loop corrections to the gauge kinetic function exhibit Mathieu moonshine. We conclude in section \ref{sec:conclusion}. Our conventions are summarized in appendix \ref{app:conventions}.

\section{The $E_8 \times E_8$ heterotic string theory compactified on toroidal orbifolds}\label{sec:setup}
We discuss compactifications of the $E_8 \times E_8$  heterotic string theory on $T^6/G$, where $G$ denotes either $\mathbb{Z}_N$ or $\mathbb{Z}_N \times \mathbb{Z}_M$. The resulting low-energy effective theory is four dimensional $\mathcal{N}=1$ supergravity. This theory consists of one gravity multiplet, a number of vector multiplets that include the vectors $A^\alpha$ and chiral multiplets with (complex) scalar components $\phi^I$. The two derivative action is completely determined by three functions: the K\"ahler potential $K$, the holomorphic superpotential $W$ and the holomorphic gauge-kinetic function $f_{\alpha\beta}$. The two derivative action is
\be
S\! = \!-\!\! \int \! \left[ -\frac12 R \star \! 1 + \! K_{I\bar{J}} d\phi^I \wedge \star d\bar{\phi}^{\bar{J}} + V \!\star \! 1 +\frac12 \text{Re}(f_{\alpha\beta}) F^\alpha \!\wedge \star F^\beta \! + \frac12 \text{Im}(f_{\alpha\beta}) F^\alpha \! \wedge F^\beta\right]\!,
\ee
where the scalar potential is given by
\be
V= e^K \lp K^{I\bar{J}} D_{\phi^I} W \overline{D_{\phi^J} W} -3|W|^2 \rp +\frac12 \text{Re} (f)^{-1\,\alpha \beta} \text{D}_\alpha \text{D}_\beta \,.
\ee
The derivatives $D_{\phi^I} W = \partial_{\phi^I} W  + W \partial_{\phi^I} K$ should not be confused with the D-terms which are
\be
\text{D}_\alpha = i \delta_\alpha \phi^I \partial_{\phi^I} K + i \frac{\delta_\alpha W}{W}\,.
\ee
Here the variation of $\phi^I$ under the infinitesimal gauge transformations $A^\alpha \rightarrow A^\alpha + \lambda^\alpha$ is $\lambda^\alpha \delta_\alpha \phi^I$ and similarly for $W$.

The derivation of the gauge group and matter content, as well as of $K$, $W$ and $f_{\alpha \beta}$, for the case of toroidal orbifold compactifications of the $E_8 \times E_8$ heterotic string have long been textbook material and we refer the reader to chapters 16 and 17 of \cite{Polchinski:1998rr} for many details and a worked out example. Here we recall a few relevant facts that will be important in the next section when we connect the 1-loop corrections of the gauge kinetic function to the moonshine phenomena discovered in \cite{EOT}.

Compactifying the heterotic string theory on $T^6$ leads to a four dimensional theory that preserves $\mathcal{N}=4$. In order to break the supersymmetry we orbifold the six torus and the $E_8 \times E_8$ gauge bundle by a discrete abelian group $G$ which we take to be either $\mathbb{Z}_N$ or $\mathbb{Z}_N \times \mathbb{Z}_M$. We define three complex coordinates on the $T^6=T^2 \times T^2 \times T^2$ by $z_j = y_{2j-1} +i U_j\, y_{2j}$, where the complex $U_j$'s denote the complex structure moduli of the $T^2$'s. We will denote the K\"ahler moduli that control the sizes of the three $T^2$'s by $T_j$ and combine the current algebra fermions into two sets of eight complex fermions which we denote by $\lambda_A$ and $\tilde{\lambda}_A$. The orbifold action for $\mathbb{Z}_N$ is then fixed by specifying the action of a generator $g$ on the spacetime coordinates\footnote{The right-moving world sheet supersymmetry fixes the action on the right-moving complex fermions to be the same as on the $z_j$.} and the gauge bundle
\be
g: z_j \rightarrow e^{\frac{2 \pi i \varphi_j}{N}} z_j\,, \qquad g: \lambda_A \rightarrow e^{\frac{2 \pi i \gamma_A}{N}} \lambda_A\,, \qquad g: \tilde\lambda_A \rightarrow e^{\frac{2 \pi i \tilde\gamma_A}{N}} \tilde \lambda_A\,.
\ee
For the case of $\mathbb{Z}_N \times \mathbb{Z}_M$ we need to specify the action of a second generator which we denote $\hat{g}$ with corresponding angles $\hat{\varphi}_j, \hat{\gamma}_A$ and $\hat{\tilde{\gamma}}_A$.

The actual values of $N$ (and $M$) are constraint by the requirement that they preserve the lattice $\Gamma$ which defines the torus $T^6 = \mathbb{R}^6/\Gamma$. Furthermore, in order to ensure that the resulting four dimensional theory preserves only $\mathcal{N}=1$ supersymmetry we restrict to groups $G$ that are contained in $SU(3)$ but not in $SU(2)$. We refer the interested reader to \cite{Reffert:2006du} for a detailed discussion of toroidal orbifolds. In tables \ref{tab:cyclicgroups} and \ref{tab:productgroups}, which are taken from \cite{Flauger:2008ad}, we list the possible orbifold actions on $T^6$ (as well as the unfixed moduli that appear in $\mathcal{N}=2$ subsectors and play a crucial role below).

\begin{table}
\begin{centering}
\begin{tabular}{|c|c|c|}
  \hline
  Group $\mathbb{Z}_N$ & Generator $\tfrac{1}{N} (\varphi_1, \varphi_2, \varphi_3)$ & $\mathcal{N}=2$ moduli\\\hline\hline
  $\mathbb{Z}_3$ & $\tfrac{1}{3} (1,1,1)$ & -\\\hline
  $\mathbb{Z}_4$ & $\tfrac{1}{4} (1,1,2)$ & $T_3, U_3$\\\hline
  $\mathbb{Z}_{6-I}$ & $\tfrac{1}{6} (1,1,4)$ & $T_3$\\\hline
  $\mathbb{Z}_{6-II}$ & $\tfrac{1}{6} (1,2,3)$ & $T_2,T_3, U_3$\\\hline
  $\mathbb{Z}_{7}$ & $\tfrac{1}{7} (1,2,4)$ & -\\\hline
  $\mathbb{Z}_{8-I}$ & $\tfrac{1}{8} (1,2,5)$ & $T_2$\\\hline
  $\mathbb{Z}_{8-II}$ & $\tfrac{1}{8} (1,3,4)$ & $T_3, U_3$\\\hline
  $\mathbb{Z}_{12-I}$ & $\tfrac{1}{12} (1,4,7)$ & $T_2$\\\hline
  $\mathbb{Z}_{12-II}$ & $\tfrac{1}{12} (1,5,6)$ & $T_3, U_3$\\  \hline
\end{tabular}
\caption{Cyclic orbifold groups.}\label{tab:cyclicgroups}
\end{centering}
\end{table}

\begin{table}
\begin{centering}
\begin{tabular}{|c|c|c|c|}
  \hline
  $\mathbb{Z}_N \times \mathbb{Z}_M$ & $\!1^{\text{st}}$ generator $\tfrac{1}{N} (\varphi_1, \varphi_2, \varphi_3)\!$ & $\!2^{\text{nd}}$ generator $\tfrac{1}{M} (\hat \varphi_1, \hat\varphi_2, \hat\varphi_3)\!$ & $\mathcal{N}=2$ moduli\\\hline\hline
  $\mathbb{Z}_2 \times \mathbb{Z}_2$ & $\tfrac{1}{2} (1,0,1)$ & $\tfrac{1}{2} (0,1,1)$ & $T_1, U_1, T_2, U_2, T_3, U_3$\\\hline
  $\mathbb{Z}_2 \times \mathbb{Z}_4$ & $\tfrac{1}{2} (1,0,1)$ & $\tfrac{1}{4} (0,1,3)$ & $T_1, U_1, T_2, T_3$\\\hline
  $\mathbb{Z}_2 \times \mathbb{Z}_6$ & $\tfrac{1}{2} (1,0,1)$ & $\tfrac{1}{6} (0,1,5)$ & $T_1, U_1, T_2, T_3$\\\hline
  $\mathbb{Z}_2 \times \mathbb{Z}_6'$ & $\tfrac{1}{2} (1,0,1)$ & $\tfrac{1}{6} (1,1,4)$ & $T_1, T_2, T_3$\\\hline
  $\mathbb{Z}_3 \times \mathbb{Z}_3$ & $\tfrac{1}{3} (1,0,2)$ & $\tfrac{1}{3} (0,1,2)$ & $T_1, T_2, T_3$\\\hline
  $\mathbb{Z}_3 \times \mathbb{Z}_6$ & $\tfrac{1}{3} (1,0,2)$ & $\tfrac{1}{6} (0,1,5)$ & $T_1, T_2, T_3$\\\hline
  $\mathbb{Z}_4 \times \mathbb{Z}_4$ & $\tfrac{1}{4} (1,0,3)$ & $\tfrac{1}{4} (0,1,3)$ & $T_1, T_2, T_3$\\\hline
  $\mathbb{Z}_6 \times \mathbb{Z}_6$ & $\tfrac{1}{6} (1,0,5)$ & $\tfrac{1}{6} (0,1,5)$ & $T_1, T_2, T_3$\\\hline
\end{tabular}
\caption{Product orbifold groups.}\label{tab:productgroups}
\end{centering}
\end{table}

The $\gamma_A$ and $\tilde{\gamma}_A$ are only defined up to shifts by $E_8$ root vectors and modulo the action of the Weyl group of $E_8$. Furthermore they have to satisfy certain constraints in order to ensure left-right level matching which guarantees modular invariance of the string path integral measure at 1-loop order \cite{Vafa:1986wx}. These conditions are
\ba\label{eq:modinv1}
\sum_{j=1}^3 (\varphi_j)^2 &=& \sum_{A=1}^8 (\gamma_A)^2 + \sum_{A=1}^8 (\tilde \gamma_A)^2\qquad \text{mod } 2N\,,\nn\\
\sum_{j=1}^3 \varphi_j &=& \sum_{A=1}^8 \gamma_A = \sum_{A=1}^8 \tilde \gamma_A = 0\qquad \text{mod } 2\,,\\
\ea
for even $N$ and
\be\label{eq:modinv2}
\sum_{j=1}^3 (\varphi_j)^2 = \sum_{A=1}^8 (\gamma_A)^2 + \sum_{A=1}^8 (\tilde \gamma_A)^2\qquad \text{mod } N\,,\\
\ee
for odd $N$. For the case of $\mathbb{Z}_N \times \mathbb{Z}_M$ we have to impose the same conditions (with $N$ replaced by $M$) for the hatted angles corresponding to the action of $\hat{g}$. It is straight forward but pretty lengthy to classify all possible actions on the current algebra fermions, so we refrain from doing so here. One simple choice is the so called standard embedding for which one chooses
\ba\label{eq:standardemd}
\gamma_A &=& \{\varphi_1,\varphi_2,\varphi_3,0,0,0,0,0\},\qquad \tilde \gamma_A = \{0,0,0,0,0,0,0,0\}\,,\\
\big( \hat{\gamma}_A &=&\{\hat{\varphi}_1,\hat{\varphi}_2,\hat{\varphi}_3,0,0,0,0,0\},\qquad \hat{\tilde{\gamma}}_A = \{0,0,0,0,0,0,0,0\} \big)\,. \nn
\ea
However, we like to stress that our results hold for arbitrary consistent choices of $\gamma_A$, $\tilde \gamma_A$ (and $\hat{\gamma}_A$, $\hat{\tilde{\gamma}}_A$).

For the standard embedding the conditions \eqref{eq:modinv1} or \eqref{eq:modinv2} are trivially satisfied and the first of the two $E_8$ gauge groups gets generically\footnote{The $U(1)^2$ factor is enhanced to $SU(3)$ for $T^6/\mathbb{Z}_3$ and to $SU(2)\times U(1)$ for $T^6/\mathbb{Z}_4$ and $T^6/\mathbb{Z}_{6-I}$.} broken to $E_6 \times U(1)^2$ while the second (hidden) $E_8$ remains unbroken. The chiral matter spectrum is model dependent but contains matter in the $\underline{27}$ of $E_6$ so the resulting four dimensional $\mathcal{N}=1$ theory closely resembles a GUT model. One can further break the $E_6$ gauge group by including Wilson lines and it is possible to get the exact chiral MSSM spectrum from certain toroidal orbifold compactifications of the heterotic string theory. However, to make the connection to Mathieu moonshine transparent, we will refrain from trying to get an $SU(3) \times SU(2) \times U(1)$ gauge group with the standard model spectrum and rather focus on the simplest toroidal orbifold models. It would be very interesting to work out the connection between Mathieu moonshine and fully realistic models.

In this paper we are mostly interested in the corrections to the gauge couplings and its dependence on neutral scalar fields. To that end we recall that we can write
\be
f_{\alpha \beta}(\phi^I) = \delta_{\alpha \beta} \, f_\alpha (\phi^I)\,,
\ee
where $f_\alpha$ is the same for all gauge bosons that belong to the same simple gauge group.

For compactifications of the heterotic string, the gauge kinetic function at tree-level is universally given by the axion-dilaton $S=e^{-2\phi}+i a$ whose real part is the (inverse) string coupling that also sets the gauge coupling. The imaginary part is the axion obtained by dualizing the $B_2$ field in four dimensions $da = \star_4 dB_2$. Due to a renormalization theorem \cite{Antoniadis199137}, the gauge kinetic function receives only perturbative corrections at 1-loop so that we have
\be
f_\alpha(\phi^I) = S + f^{\text{1-loop}}_\alpha(\phi^I) + \mathcal{O}(e^{-2\pi S})\,.
\ee
The function $f^{\text{1-loop}}_\alpha(\phi^I)$ is the key player in this paper. We show in the next section that whenever it has a non-trivial dependence on the moduli, then it can be expanded in such a way that the expansion coefficients are dimensions of representations of M$_{24}$.

Before we do so we recall several facts about the contributions to $f^{\text{1-loop}}_\alpha(\phi^I)$ from \cite{Dixon1991649}. To that end it is useful to introduce the concept of `different subsectors' of the orbifold compactification: We say that the orbifold compactification has an $\mathcal{N}=2$ subsector whenever there exists a non-trivial subgroup $G' \subset G$ such that the compactification $T^6/G'$ preserves $\mathcal{N}=2$ supersymmetry in four dimensions. This is the case whenever $G' \subset SU(2)$ in which case $T^6/G' = T^2 \times T^4/G'$, with $T^4/G'$ being an orbifold limit of a $K3$ manifold. For example, for $T^6/\bZ_4$ the generator $g$ of $\bZ_4 = \{ 1, g, g^2 ,g^3\}$ acts on the three complex coordinates as $g:(z_1,z_2,z_3) \rightarrow (i\,z_1,i\,z_2,-z_3)$. The $\bZ_2$ subgroup $G'=\{1,g^2\}$ does not act on $z_3$ and therefore leads to an $\N=2$ subsector. Looking at tables \ref{tab:cyclicgroups} and \ref{tab:productgroups} we see that such subsectors exist for $\mathbb{Z}_N$ whenever $N\neq 3,7$ and for all $\mathbb{Z}_N\times \mathbb{Z}_M$ orbifold models. Actually most of the models have multiple $\N=2$ subsectors which we label by $G'_j$, where the $j$ subscript means that the $j$-th $T^2$ with coordinates $z_j$ is fixed under $G'_j$. In the case that there is no non-trivial $G'_j$ for a particular $j$, we take $G'_j=\{\}$ to be the empty group. Tables \ref{tab:cyclicgroups} and \ref{tab:productgroups} list in the last column the unfixed moduli of the $j$-th $T^2$ whenever there exists a non-trivial $G'_j$. For example for $\bZ_2 \times \bZ_6$ which is generated by $g$ and $\hat{g}$, we have $G'_1 = \bZ_6 = <g'>$. The corresponding moduli of the first $T^2$ are $T_1$ and $U_1$ and they are both moduli of the full orbifold $T^6/\bZ_2 \times \bZ_6$. Therefore they both appear in the last column of table \ref{tab:productgroups}. $G'_2 = \bZ_2$ is generated by $g$ and the moduli of the second $T^2$ are $T_2$ and $U_2$. However, the full $T^6/\bZ_2 \times \bZ_6$ orbifold fixes $U_2 = e^{\pi i/6}$ so that it does not appear in the last column. Lastly we have $G'_3 = \bZ_2 = \{1, g \hat{g}^3 \}$ and $U_3 = e^{\pi i/6}$ is again fixed.

Similarly to the $\N=2$ sectors, all orbifold compactifications have an $\mathcal{N}=4$ subsector which is just the untwisted sector that arises from the compactification on $T^6$ (i.e. for $G'=\{1\}$ being the trivial group).

The gauge coupling is not renormalized in theories that preserve $\mathcal{N}=4$ supersymmetry, so $\mathcal{N}=4$ subsectors do not contribute to the gauge kinetic function. In our setup we give a simple argument for this below. Dixon, Louis and Kaplunovsky \cite{Dixon1991649} have furthermore shown that, for toroidal orbifold compactifications of the heterotic string, the only moduli dependent corrections to $f^{\text{1-loop}}_\alpha(\phi^I)$ arise from $\mathcal{N}=2$ subsectors.\footnote{Their argument, which we recall below in section \ref{sec:main}, only applies to the dependence on untwisted moduli.} In particular (up to a constant) they are given by (cf. for example \cite{Dixon1991649, Louistwo, Kiritsis, Stieberger})
\ba\label{eq:1loop}
f^{\text{1-loop}}_\alpha(T_j,U_j) =\sum_{j=1,2,3}\frac{|G'_j|}{|G|} &&\left[-
\frac12 \partial_{T_j} \partial_{U_j} h^{1-loop}_j(T_j,U_j)-\frac{1}{8\pi^2} \log\lp J(q_{T_j})-J(q_{U_j}) \rp \right.\cr
&&\left.-\frac{b_{\a,j}^{(\N=2)}}{4 \pi^2} \lp \log(\eta(q_{T_j})) + \log(\eta(q_{U_j})) \rp \right]\,,
\ea
where we used $q_{T_j}=e^{-2\pi T_j}$ and $q_{U_j}=e^{-2\pi U_j}$. $h^{1-loop}_j$ denotes the 1-loop correction to the $\N=2$ prepotential of the $\N=2$ theory obtained by a compactification on $T^2 \times T^4/G_j'$ and $b_{\a,j}^{(\N=2)}$ is the corresponding beta function. Note that \eqref{eq:1loop} implies that $T^6/\bZ_3$ and $T^6/\bZ_7$ orbifolds have gauge kinetic functions that are (perturbatively) exact at tree-level, since these orbifolds have no $\N=2$ subsectors. These are the only orbifolds with gauge kinetic functions that are not related to M$_{24}$.

In the next section, we work out the prepotential for four dimensional $\N=2$ compactifications, connect it to Mathieu moonshine and then use \eqref{eq:1loop} to show how the gauge kinetic functions in $\N=1$ toroidal compactifications are related to M$_{24}$. Before we do so, it might be illuminating to explicitly spell out \eqref{eq:1loop} for one explicit case. Let us consider $T^6/\bZ_{6-II}$ for which we have $G'_1=\{\}$, $G'_2 = \bZ_2 = \{1,g^3\}$ and $G'_3 = \bZ_3 = \{1,g^2,g^4\}$. This leads to
\ba
f^{\text{1-loop}}_\alpha(T_2,T_3,U_3) = \frac{2}{6} && \left[-\frac12 \partial_{T_2} \partial_{U_2} h^{1-loop}_2(T_2,U_2)-\frac{1}{8\pi^2} \log\lp J(q_{T_2})-J(q_{U_2}) \rp \right.\cr
&&\left.-\frac{b_{\a,2}^{(\N=2)}}{4 \pi^2} \lp \log(\eta(q_{T_2})) + \log(\eta(q_{U_2})) \rp \right]_{U_2 = e^{\frac{\pi i}{6}}} \cr
+\frac{3}{6} &&\left[-\frac12 \partial_{T_3} \partial_{U_3} h^{1-loop}_3 (T_3,U_3)-\frac{1}{8\pi^2} \log\lp J(q_{T_3})-J(q_{U_3}) \rp \right.\cr
&&\left.-\frac{b_{\a,3}^{(\N=2)}}{4 \pi^2} \lp \log(\eta(q_{T_3})) + \log(\eta(q_{U_3})) \rp \right]\,.
\ea
Although the modulus $T_1$ is not fixed, it does not appear in $f^{\text{1-loop}}_\alpha$ since $G'_1=\{\}$. Furthermore, the $U_2$ modulus is fixed by the $\bZ_{6-II}$ action to be $U_2 = e^{\pi i /6}$. $h^{1-loop}_2(T_2,U_2)$ and $b_{\a,2}^{(\N=2)}$ are determined by a compactification on $T^2 \times T^4/\bZ_2$ were the action on the $E_8\times E_8$ fermions is given by $3 \gamma_A$ and $3 \tilde{\gamma}_A$ and $T_2$ and $U_2$ are the moduli of the $T^2$ factor in $T^2 \times T^4/\bZ_2$. Similarly, $h^{1-loop}_3(T_3,U_3)$ and $b_{\a,3}^{(\N=2)}$ are determined by a compactification on $T^2 \times T^4/\bZ_3$ were the action on the $E_8\times E_8$ fermions is given by $2 \gamma_A$ and $2 \tilde{\gamma}_A$ and $T_3$ and $U_3$ are the moduli of the $T^2$ factor in $T^2 \times T^4/\bZ_3$.

\section{1-loop threshold corrections and moonshine}\label{sec:main}
As we have seen in the previous section, the moduli dependent part of the 1-loop corrections to the gauge kinetic functions arises entirely from subsectors that preserve $\mathcal{N}=2$ spacetime supersymmetry. Thus we review the general form of the threshold corrections in four dimensional $\N=2$ theories obtained from compactifications on $T^2 \times K3$ (cf. for example \cite{Harvey-Moore, Stieberger}).

\subsection {Threshold corrections in four-dimensional $\N=2$ theories}
For compactifications of the heterotic string on $T^2 \times T^4/G'$, the 1-loop string threshold correction $\Delta_\a$ for the $\a$-th gauge group is given by \cite{Antoniadis}
\be\label{eq:Delta}
\Delta_\a = \int_\mathcal{F} \frac{d^2\tau}{\tau_2} \ls -\frac{i}{\eta(q)^2} \text{Tr}_R \lp J_0 e^{i \pi J_0} q^{L_0-\frac{c}{24}}\bar{q}^{\bar{L}_0-\frac{\bar{c}}{24}} \lp Q_\a^2 -\frac{1}{8\pi \tau_2} \rp \rp -b_\a^{(\N=2)}\rs\,.
\ee
Here the trace is taken over all left-moving boundary conditions but only over the right moving Ramond sector of the $(c,\bar{c}) = (22,9)$ internal CFT theory associated with the toroidal orbifold and the left-moving $E_8 \times E_8$ gauge bundle. We use convention for which $q=e^{2 \pi i \tau}$ with $\tau$ the complex structure modulus of the 1-loop string worldsheet and $\tau_2$ its imaginary part. The prefactor $1/\eta(q)^2$ arises from the two additional four dimensional spacetime bosons in lightcone gauge. $Q_\a$ denotes the gauge charge and $b_\a^{(\N=2)}$ the 1-loop beta function.

The integrand is essentially the new supersymmetric index \cite{Vafa}
\be
\mathcal{Z}_{new} = \frac{1}{\eta(q)^2} \text{Tr}_R \lp J_0 e^{i \pi J_0} q^{L_0-\frac{c}{24}}\bar{q}^{\bar{L}_0-\frac{\bar{c}}{24}} \rp\,,
\ee
weighted by the gauge charge squared. It was shown in \cite{Harvey-Moore} that the new supersymmetric index counts BPS states in four dimensional $\mathcal{N}=2$ theories. We have schematically
\be\label{eq:ZnewBPS}
\mathcal{Z}_{new} = -2i \ls \sum_{BPS\; vectors} q^n \bar{q}^{\bar{n}} - \sum_{BPS\; hypers} q^n \bar{q}^{\bar{n}} \rs\,,
\ee
which shows that a subsector which preserves $\mathcal{N}=4$ does not contribute since $\mathcal{N}=4$ BPS states split into one $\mathcal{N}=2$ hypermultiplet and one $\mathcal{N}=2$ vectormultiplet, which then cancel each other in \eqref{eq:ZnewBPS}.

Furthermore, it was shown in \cite{Harvey-Moore} that for compactifications on $T^2 \times T^4/G'$ the new supersymmetric index is closely related to the elliptic genus
\be
\mathcal{Z}_{elliptic}(q,y) = \text{Tr}_{RR} \lp (-1)^{F_L + F_R} q^{L_0-\frac{c}{24}} y^{J_0}\bar{q}^{\bar{L}_0-\frac{\bar{c}}{24}} \rp\,.
\ee
The trace in the elliptic genus is taken over the left- and right-moving Ramond sectors, $F_{L/R}$ are the fermion number operators and $y = e^{2 \pi i z}$ a chemical potential for the $U(1)$-charge measured by $J_0$. One finds that \cite{Harvey-Moore, Cheng:2013kpa} \footnote{Please see appendix \ref{app:conventions} for our conventions.}
\ba\label{eq:ZnewZK3}
\mathcal{Z}_{new}&=& \frac{i}{2} \frac{\Theta_{\Gamma_{2,2}}(q, \bar{q}; T, U, \bar{T},\bar{U}) E_4(q)}{\eta(q)^{12}} \Bigg[ \lp \frac{\theta_2(q)}{\eta(q)} \rp^6 \cZ_{elliptic}^{K3}(q,-1) \cr
&& +q^{\frac14} \lp \frac{\theta_3(q)}{\eta(q)} \rp^6 \cZ_{elliptic}^{K3}(q,-\sqrt{q}) -q^{\frac14} \lp \frac{\theta_4(q)}{\eta(q)} \rp^6 \cZ_{elliptic}^{K3}(q,\sqrt{q}) \Bigg]\,,
\ea
where
\be\label{eq:ZK3}
\cZ_{elliptic}^{K3}(q,y) = 8 \left[ \lp \frac{\theta_2(q,y)}{\theta_2(q,1)} \rp^2 +\lp \frac{\theta_3(q,y)}{\theta_3(q,1)} \rp^2 +\lp \frac{\theta_4(q,y)}{\theta_4(q,1)} \rp^2 \right]\,,
\ee
is the elliptic genus of $K3$ and 
\be
\Theta_{\Gamma_{2,2}}= \sum_{p \in \Gamma_{2,2}} q^{\frac12 p_L^2} \bar{q}^{\frac12 p_R^2} = \sum_{m_i,n_i \in \bZ} e^{2\pi i \tau (m_1 n_1 + m_2 n_2) -\frac{\pi \tau_2}{\text{Re}(T)\text{Re}(U)} |-T U n_2 + i T n_1- i U m_1 + m_2 |}\,,
\ee
is the sum over windings and momenta on the $T^2$. $\Theta_{\Gamma_{2,2}}$ is the only contribution that depends on the moduli of the toroidal orbifold and such a dependence can only arise in $\N=2$ subsectors. This can be nicely seen from \eqref{eq:Delta}. The only dependence on the untwisted moduli arises from $L_0$ and $\bar{L}_0$ for states with non-trivial momenta and/or winding numbers. For toroidal orbifolds we have to sum over all different boundary conditions, twisted by $(g,h)$ along the two cycles of the string world-sheet. The only boundary conditions for which the trace receives contributions from windings and momenta are such that $(g,h)$ both do not act on a $T^2$ factor, i.e. for $\N=2$ subsectors.\footnote{Recall that $G \subset SU(3)$ contains no non-trivial element that preserves a $T^4$. There could be also moduli dependent contributions when $(g,h)=(1,1)$ but as argued above this $\N=4$ subsector gives a vanishing contribution.}

The reason that $\cZ_{elliptic}^{K3}(q,y)$ appears in \eqref{eq:ZnewZK3} for all compactifications on $T^2 \times T^4/G'$ can be understood by the fact that $\cZ_{elliptic}^{K3}(q,y)$ is an index an therefore does not change when moving in $K3$ moduli space, even when going to the orbifold limit $T^4/G'$. The answer for $\mathcal{Z}_{new}$ is also the same for all possible orbifold actions on the current algebra fermions (i.e. for all different $\gamma_A$'s). These choices determine how we embed instantons into the $E_8 \times E_8$ gauge group. The Bianchi identity for $H_3 = dB_2$ requires us to embed a total of 24 instantons into $E_8 \times E_8$ and we denote the number of instantons in the two $E_8$'s by $(n_1,n_2)$. For a supersymmetry preserving compactification we have to demand that $n_{1}, n_2 \geq 0$ in addition to $n_1 + n_2 =24$.\footnote{For example the standard embedding in \eqref{eq:standardemd} has $(n_1,n_2)=(24,0)$.} The transitions between models with different instanton numbers is non-perturbative, so the invariance of $\mathcal{Z}_{new}$ cannot be explained by the fact that it is an index. However, one can use the transformation properties of $\mathcal{Z}_{new}$ under $SL(2;\bZ)$ together with its pole structure to argue that it has to be uniquely given by \eqref{eq:ZnewZK3} (cf. \cite{Kiritsis, Moore, Cheng:2013kpa}).

As was shown in \cite{EOT}, the elliptic genus of $K3$ can be expanded in terms of $\N=4$ Virasoro characters
\ba
\text{ch}_{h=\frac14,l=0}(q,y) &=& -\frac{i \sqrt{y} \theta_1(q,y)}{\eta(q)^3} \sum_{n=- \infty}^\infty \frac{(-1)^n q^{\frac12 n(n+1)} y^n}{1-q^n y}\,,\cr
\text{ch}_{h=n+\frac14,l=\frac12}(q,y) &=& q^{n-\frac18} \frac{\theta_1(q,y)^2}{\eta(q)^3}\,,
\ea
in such a way that first few expansion coefficients are positive sums of irreps of M$_{24}$
\be
\cZ_{elliptic}^{K3}(q,y) = 24 \text{ch}_{h=\frac14,l=0}(q,y) -2\text{ch}_{h=\frac14,l=\frac12}(q,y) + \sum_{n=1}^\infty A_n \text{ch}_{h=n+\frac14,l=\frac12}(q,y)\,.
\ee
In particular, one finds
\be\label{eq:An}
A_1 = 45+\underline{45},\quad A_2 = 231+\underline{231},\quad A_3 = 770+\underline{770},\quad A_4 = 2 \cdot 2277, \quad \ldots
\ee
where for example $45$ and $\underline{45}$ denote the two different 45 dimensional irreps of M$_{24}$.

Several non-trivial checks that confirmed the connection between M$_{24}$ and $\cZ_{elliptic}^{K3}$ were performed in \cite{Miranda, Gaberdiel:2010ch, Gaberdiel:2010ca, Eguchi:2010fg}. In \cite{Gannonproof} it was then shown that all the $A_n$ are positive sums of dimension of irreps of M$_{24}$, which provides very strong evidence for a moonshine phenomena relating M$_{24}$ and $K3$. This connection is however currently not understood. In particular, it was shown in \cite{Gaberdiel} that no $\N=(4,4)$ sigma model with $K3$ target space has M$_{24}$ as its symmetry group.

Plugging \eqref{eq:ZK3} into \eqref{eq:ZnewZK3} one finds
\be
\mathcal{Z}_{new}= -2i \Theta_{\Gamma_{2,2}} \frac{E_4(q) E_6(q)}{\eta(q)^{24}}\,.
\ee
Defining
\ba\label{eq:gs}
g_{h=\frac14,l}(q) &=& \lp \frac{\theta_2(q)}{\eta(q)} \rp^6 \text{ch}_{h=\frac14,l}(q,-1) + q^{\frac14} \lp \frac{\theta_3(q)}{\eta(q)} \rp^6 \text{ch}_{h=\frac14,l}(q,-\sqrt{q}) \cr
&&- q^{\frac14} \lp \frac{\theta_4(q)}{\eta(q)} \rp^6 \text{ch}_{h=\frac14,l}(q,\sqrt{q}) \,,
\ea
we can write
\be\label{eq:ZnewM24}
\mathcal{Z}_{new}= \frac{i \Theta_{\Gamma_{2,2}} E_4(q)}{2 \eta(q)^{12}} \left[ 24 g_{h=\frac14,l=0}(q) + g_{h=\frac14,l=\frac12}(q)\lp -2 +\sum_{n=1}^\infty A_n q^n \rp \right]\,,
\ee
where the $A_n$ are again given by \eqref{eq:An}. For the particular case of the standard embedding the left-moving sector has $\N=4$ world-sheet supersymmetry and the relation between $\mathcal{Z}_{new}$ and $\mathcal{Z}_{elliptic}(q,y)$ is such that the above coefficients are literally the same as in the original Mathieu moonshine observation by Eguchi, Ooguri and Tachikawa \cite{EOT}. However, there is strong evidence that even for arbitrary instanton embeddings with $\N=(0,4)$ world sheet supersymmetry the above coefficients are related to M$_{24}$ (or at least a subgroup thereof) \cite{Harrison:2013bya, 04paper}.

Having established the connection between the worldsheet quantity $\mathcal{Z}_{new}$ and the Mathieu group M$_{24}$, we are now going to connect the expansion coefficients $c(m)$ of
\be
\mathcal{Z}_{new}= -2i \Theta_{\Gamma_{2,2}} \frac{E_4(q) E_6(q)}{\eta(q)^{24}} = -2i \Theta_{\Gamma_{2,2}} \lp \sum_{m=-1}^\infty c(m) q^m \rp\,,
\ee
to a spacetime quantity, namely the prepotential that determines the vector multiplet sector of the four dimensional $\N=2$ spacetime theory. This then connects M$_{24}$ to the prepotential since we have from \eqref{eq:ZnewM24}
\be\label{eq:cm}
\sum_{m=-1}^\infty c(m) q^m = -\frac{E_4(q)}{4 \eta(q)^{12}} \left[ 24 g_{h=\frac14,l=0}(q) + g_{h=\frac14,l=\frac12}(q)\lp -2 +\sum_{n=1}^\infty A_n q^n \rp \right]\,.
\ee
We like to mention that the equation \eqref{eq:cm} that connects the $c(m)$ to M$_{24}$ seems somewhat convoluted. This can be explained (at least for the standard embedding) by the fact that the $\N=2$ theory obtained by compactifications on $T^2 \times K3$ has $E_7\times E_8$ gauge group in addition to the M$_{24}$ moonshine. In particular the $E_8$ gauge group leads to a factor $E_4(q)/\eta(q)^8$ in \eqref{eq:cm}. While the $E_7$ is not manifest, we have (for the standard embedding) from the left moving free fermions an affine $SO(12)$ current algebra. This explains the $\theta_i(q)^6/\eta(q)^6$ in front of the $\N=4$ Virasoro character in the definition of the $g_{h=\frac14,l}(q)$ in \eqref{eq:gs}. It is thus natural to decompose the $c(m)$ into representations of $E_8 \times SO(12) \times M_{24}$. While this conclusion seems inevitable for the standard embedding, it is not so clear for arbitrary instanton embeddings. For these more generic cases there is clear evidence for a connection to (at least a subgroup of) M$_{24}$ \cite{Harrison:2013bya}, however the $E_8 \times SO(12)$ symmetry is generically broken and it is unclear whether it is miraculous restored in $\mathcal{Z}_{new}$ or not (see \cite{04paper} for further results).

To connect the $c(m)$ in \eqref{eq:cm} to the prepotential one has to perform the integral \eqref{eq:Delta}, which then determines the 1-loop correction to the $\N=2$ prepotential \cite{Harvey-Moore, Cheng:2013kpa} to be
\be\label{eq:h1loop}
h^{1-loop}(T,U) = -\frac{1}{12\pi} U^3 -\frac{1}{2(2 \pi)^4} c(0) \zeta(3)-\frac{1}{(2 \pi)^4} \sum_{\substack{k>0, l \in \bZ \\ k=0,l>0}} c(kl) Li_3 \lp e^{-2\pi(k T+ l U)}\rp\,.
\ee
Here $T$ and $U$ are the K\"ahler and complex structure moduli of the $T^2$ in $T^2 \times T^4/G'$ and $Li_3(x) = \sum_{n=1}^\infty \tfrac{x^n}{n^3}$. The coefficients $c(m)$ are exactly the same as the ones in the expansion \eqref{eq:cm} and thus still related to M$_{24}$. The full $\N=2$ prepotential receives only perturbative corrections at one loop and is given by
\be
h(T,U) = -STU+h^{1-loop} + \mathcal{O}(e^{-2\pi S})\,.
\ee
Lastly we note that the $\N=2$ beta functions for a particular instanton embedding $(n_1,n_2)$ that breaks $E_8 \times E_8$ to $H_1 \times H_2$ are given by \cite{Stieberger}
\be\label{eq:N=2beta}
b_{H_1}^{(\N=2)} = -60+6n_1\,,\qquad b_{H_2}^{(\N=2)} = -60+6n_2\,.
\ee
This concludes our excursion into compactifications of the heterotic string theory that preserve $\N=2$ supersymmetry in four dimension. We can now combine the results above with the previous section to connect the gauge coupling in our four dimensional $\N=1$ theories to the sporadic group M$_{24}$.

\subsection{Mathieu moonshine in $\N=1$ theories}
We now use the results above in the equation for the 1-loop correction to the gauge kinetic function \eqref{eq:1loop} which he repeat here
\ba\label{eq:1-loop2}
f^{\text{1-loop}}_\alpha(T_j,U_j) =\sum_{j=1,2,3}\frac{|G'_j|}{|G|} &&\left[-\frac12 \partial_{T_j} \partial_{U_j} h^{1-loop}_j(T_j,U_j)-\frac{1}{8\pi^2} \log\lp J(q_{T_j})-J(q_{U_j}) \rp \right.\cr
&&\left.-\frac{b_{\a,j}^{(\N=2)}}{4 \pi^2} \lp \log(\eta(q_{T_j})) + \log(\eta(q_{U_j})) \rp \right]\,.
\ea
To get a more explicit expression we use the result \eqref{eq:h1loop} to find
\ba
\partial_{T_j} \partial_{U_j} h^{1-loop}_j(T_j,U_j) &=& \partial_{T_j} \partial_{U_j} \Bigg( -\frac{1}{(2 \pi)^4} \sum_{\substack{k>0, l \in \bZ \\ k=0,l>0}} c(kl) Li_3 \lp e^{-2\pi(k T_j+ l U_j)} \rp \Bigg) \cr
&=& -\frac{1}{(2 \pi)^4} \partial_{T_j} \partial_{U_j} \Bigg( \sum_{\substack{k=1,l=-1 \\ k,l>0}} c(kl) Li_3 \lp e^{-2\pi(k T_j+ l U_j)} \rp \Bigg) \cr
&=& \frac{1}{(2 \pi)^2} \Bigg(-\log \lp 1-\frac{q_{T_j}}{q_{U_j}} \rp +\sum_{k,l >0} c(kl) k l \log \lp 1-q_{T_j}^k q_{U_j}^l \rp \Bigg)\,,\qquad \quad
\ea
where we used that $c(m)=0$ for $m \leq -2$ and $c(-1)=1$. We can now rewrite the 1-loop correction as the sum over a logarithm
\ba
f^{\text{1-loop}}_\alpha(T_j,U_j) =\sum_{j=1,2,3}\frac{|G'_j|}{8 \pi^2|G|} \log \lp\frac{\lp1-\frac{q_{T_j}}{q_{U_j}}\rp \lp \eta(q_{T_j})\eta(q_{U_j})\rp^{-2 b_{\a,j}^{(\N=2)}}}{\lp J(q_{T_j})-J(q_{U_j})\rp\prod_{k,l >0} \lp 1-q_{T_j}^k q_{U_j}^l \rp^{c(kl) k l}} \rp \,,\quad
\ea
however, the different terms in \eqref{eq:1-loop2} make the contributions to the 1-loop correction for the gauge kinetic functions somewhat more transparent and make it clear that the first term is directly related to the $\N=2$ prepotential and is thus related to M$_{24}$.

\section{Conclusion}\label{sec:conclusion}
In this paper we have shown that the moduli dependence of the 1-loop corrected gauge kinetic function in toroidal compactifications of the heterotic string is connected to the sporadic group M$_{24}$. The starting point that allowed us to make this connection is the discovery made by Eguchi, Ooguri, and Tachikawa \cite{EOT} that connects $K3$ manifolds with the largest Mathieu group M$_{24}$. One can show that this implies that string threshold corrections in heterotic compactifications on $T^2 \times K3$ are also related to M$_{24}$. Using the fact that the gauge kinetic function in heterotic toroidal orbifold compactifications only receives moduli dependent corrections from $\N=2$ subsectors, we have been able to present a large class of interesting four dimensional theories with $\N=1$ supersymmetry whose 1-loop corrections to the gauge couplings are controlled by the Mathieu group M$_{24}$.

While our class of models contains GUT-like theories, it would be interesting to study whether it is possible to obtain the MSSM while still preserving the connection to M$_{24}$. It would also be interesting to investigate whether the slightly involved expansion of the gauge kinetic function, that is needed in order to extract the M$_{24}$ coefficients, hints at the presence of a larger (sporadic) group. A few steps towards understanding such a larger structure are taken in \cite{04paper}.

Using string duality it was shown \cite{Cheng:2013kpa} that Mathieu moonshine implies that $CY_3$ manifolds that are elliptically fibered over a Hirzebruch surface $\mathbb{F}_n$ have Gromov-Witten invariants that are likewise connected to M$_{24}$. Furthermore, in \cite{CY4paper} several new connections between $CY_4$ manifolds and sporadic groups are presented. Since $K3$, $CY_3$ and $CY_4$ manifolds have been key components in string compactifications for decades, it is likely that as a consequence there are a variety of further discoveries that await us and that will connect spacetime physics and the moonshine phenomena. In particular, F-theory compactifications on elliptically fibered $CY_4$ manifolds lead to four dimensional $\N=1$ theories, so the results of \cite{CY4paper} are presumable also relevant for four dimensional spacetime physics with $\N=1$ supersymmetry. Since these F-theory compactification are often dual to the heterotic constructions discussed here, there should be a variety of interconnections that await our discovery.

\acknowledgments

I would like to thank M.~Cheng, X.~Dong, J.~Duncan, S.~Harrison, J.~Harvey, S.~Kachru, S.~Stieberger and D.~Whalen for illuminating discussions and S.~Kachru and X.~Dong for comments on the manuscript. This work was supported by a Research Fellowship (Grant number WR 166/1-1) of the German Research Foundation (DFG).

\appendix

\section{Conventions}\label{app:conventions}
Our conventions for the Jacobi functions $\theta_i(q,y)$ are
\ba
\theta_1(q,y) &=& i\!\! \sum_{n=-\infty}^{\infty}\! (-1)^n q^{\frac{(n-\frac12)^2}{2}} y^{n-\frac12} = -i q^{1/8} y^{1/2} \prod_{n=1}^{\infty} (1-q^n)(1-y q^n)(1-y^{-1}q^{n-1})\,,\qquad\quad\\
\theta_2(q,y) &=& \,\sum_{n=-\infty}^{\infty} q^{\frac{(n-\frac12)^2}{2}} y^{n-\frac12} \qquad \,\,\,= \quad q^{1/8} y^{1/2} \prod_{n=1}^{\infty} (1-q^n)(1+y q^n)(1+y^{-1}q^{n-1})\,,\\
\theta_3(q,y) &=& \,\sum_{n=-\infty}^{\infty} q^{\frac{n^2}{2}} y^n\qquad \quad \qquad \,\,= \qquad \qquad \,\,\, \prod_{n=1}^{\infty} (1-q^n)(1+y q^{n-\frac12})(1+y^{-1}q^{n-\frac12}),\\
\theta_4(q,y) &=& \,\sum_{n=-\infty}^{\infty} (-1)^n q^{\frac{n^2}{2}} y^n \qquad \quad = \qquad \qquad \,\,\, \prod_{n=1}^{\infty} (1-q^n)(1-y q^{n-\frac12})(1-y^{-1}q^{n-\frac12}).
\ea
Whenever we do not specify the $y$-dependence, we have set $y=1$, so that for example $\theta_i(q)=\theta_i(q,1)$. We use the standard definition for the Dedekind $\eta(q)$ function
\be
\eta(q) = q^{\frac{1}{24}} \sum_{n=-\infty}^{\infty} (-1)^n q^{\frac{n(3n-1)}{2}} = q^{\frac{1}{24}} \prod_{n=1}^{\infty} (1-q^n)\,.
\ee
It is also convenient to use the Eisenstein series $E_4(q)$ and $E_6(q)$ that can be written in terms of the $\theta_i(q)$ as follows
\ba
E_4(q) &=& \frac12 \lp \theta_2(q)^8 +\theta_3(q)^8 +\theta_4(q)^8\rp\,,\\
E_6(q) &=& - \frac12 \lp \theta_2(q)^8 (\theta_3(q)^4+\theta_4(q)^4) +\theta_3(q)^8 (\theta_2(q)^4-\theta_4(q)^4) -\theta_4(q)^8 (\theta_2(q)^4+\theta_3(q)^4)\rp\,.\nn\qquad
\ea
Lastly we define Klein's $J$-function
\be
J(q) = \frac{E_6(q)^2}{\eta(q)^{24}}+984 = \frac{1}{q} + 196884 q + 21493760 q^2 +\ldots
\ee

\bibliographystyle{JHEP}
\bibliography{refs}

\providecommand{\href}[2]{#2}\begingroup\raggedright\begin{thebibliography}{10}

\bibitem{EOT}
T.~Eguchi, H.~Ooguri, and Y.~Tachikawa, {\it {Notes on the K3 Surface and the
  Mathieu group $M_{24}$}},  {\em Exper.Math.} {\bf 20} (2011) 91--96,
  [\href{http://xxx.lanl.gov/abs/1004.0956}{{\tt arXiv:1004.0956}}].

\bibitem{Miranda}
M.~C. Cheng, {\it {K3 Surfaces, N=4 Dyons, and the Mathieu Group M24}},  {\em
  Commun.Num.Theor.Phys.} {\bf 4} (2010) 623--658,
  [\href{http://xxx.lanl.gov/abs/1005.5415}{{\tt arXiv:1005.5415}}].

\bibitem{Gaberdiel:2010ch}
M.~R. Gaberdiel, S.~Hohenegger, and R.~Volpato, {\it {Mathieu twining
  characters for K3}},  {\em JHEP} {\bf 1009} (2010) 058,
  [\href{http://xxx.lanl.gov/abs/1006.0221}{{\tt arXiv:1006.0221}}].

\bibitem{Gaberdiel:2010ca}
M.~R. Gaberdiel, S.~Hohenegger, and R.~Volpato, {\it {Mathieu Moonshine in the
  elliptic genus of K3}},  {\em JHEP} {\bf 1010} (2010) 062,
  [\href{http://xxx.lanl.gov/abs/1008.3778}{{\tt arXiv:1008.3778}}].

\bibitem{Eguchi:2010fg}
T.~Eguchi and K.~Hikami, {\it {Note on Twisted Elliptic Genus of K3 Surface}},
  {\em Phys.Lett.} {\bf B694} (2011) 446--455,
  [\href{http://xxx.lanl.gov/abs/1008.4924}{{\tt arXiv:1008.4924}}].

\bibitem{Gannonproof}
T.~Gannon, {\it {Much ado about Mathieu}},
  \href{http://xxx.lanl.gov/abs/1211.5531}{{\tt arXiv:1211.5531}}.

\bibitem{Hohenegger:2011us}
S.~Hohenegger and S.~Stieberger, {\it {BPS Saturated String Amplitudes: K3
  Elliptic Genus and Igusa Cusp Form}},  {\em Nucl.Phys.} {\bf B856} (2012)
  413--448, [\href{http://xxx.lanl.gov/abs/1108.0323}{{\tt arXiv:1108.0323}}].

\bibitem{Harvey:2013mda}
J.~A. Harvey and S.~Murthy, {\it {Moonshine in Fivebrane Spacetimes}},  {\em
  JHEP} {\bf 1401} (2014) 146, [\href{http://xxx.lanl.gov/abs/1307.7717}{{\tt
  arXiv:1307.7717}}].

\bibitem{Cheng:2013kpa}
M.~C. Cheng, X.~Dong, J.~Duncan, J.~Harvey, S.~Kachru, and T.~Wrase, {\it
  {Mathieu Moonshine and N=2 String Compactifications}},  {\em JHEP} {\bf 1309}
  (2013) 030, [\href{http://xxx.lanl.gov/abs/1306.4981}{{\tt
  arXiv:1306.4981}}].

\bibitem{Harvey-Moore}
J.~A. Harvey and G.~W. Moore, {\it {Algebras, BPS states, and strings}},  {\em
  Nucl.Phys.} {\bf B463} (1996) 315--368,
  [\href{http://xxx.lanl.gov/abs/hep-th/9510182}{{\tt hep-th/9510182}}].

\bibitem{Dixon1991649}
L.~J. Dixon, V.~S. Kaplunovsky, and J.~Louis, {\it Moduli dependence of string
  loop corrections to gauge coupling constants},  {\em Nuclear Physics B} {\bf
  355} (1991), no.~3 649 -- 688.

\bibitem{Polchinski:1998rr}
J.~Polchinski, {\it {String theory. Vol. 2: Superstring theory and beyond}}, .

\bibitem{Reffert:2006du}
S.~Reffert, {\it {Toroidal Orbifolds: Resolutions, Orientifolds and
  Applications in String Phenomenology}},
  \href{http://xxx.lanl.gov/abs/hep-th/0609040}{{\tt hep-th/0609040}}.

\bibitem{Flauger:2008ad}
R.~Flauger, S.~Paban, D.~Robbins, and T.~Wrase, {\it {Searching for slow-roll
  moduli inflation in massive type IIA supergravity with metric fluxes}},  {\em
  Phys.Rev.} {\bf D79} (2009) 086011,
  [\href{http://xxx.lanl.gov/abs/0812.3886}{{\tt arXiv:0812.3886}}].

\bibitem{Vafa:1986wx}
C.~Vafa, {\it {Modular Invariance and Discrete Torsion on Orbifolds}},  {\em
  Nucl.Phys.} {\bf B273} (1986) 592.

\bibitem{Antoniadis199137}
I.~Antoniadis, K.~Narain, and T.~Taylor, {\it Higher genus string corrections
  to gauge couplings},  {\em Physics Letters B} {\bf 267} (1991), no.~1 37 --
  45.

\bibitem{Louistwo}
B.~de~Wit, V.~Kaplunovsky, J.~Louis, and D.~L{\"u}st, {\it {Perturbative
  couplings of vector multiplets in N=2 heterotic string vacua}},  {\em
  Nucl.Phys.} {\bf B451} (1995) 53--95,
  [\href{http://xxx.lanl.gov/abs/hep-th/9504006}{{\tt hep-th/9504006}}].

\bibitem{Kiritsis}
E.~Kiritsis, C.~Kounnas, P.~Petropoulos, and J.~Rizos, {\it {Universality
  properties of N=2 and N=1 heterotic threshold corrections}},  {\em
  Nucl.Phys.} {\bf B483} (1997) 141--171,
  [\href{http://xxx.lanl.gov/abs/hep-th/9608034}{{\tt hep-th/9608034}}].

\bibitem{Stieberger}
S.~Stieberger, {\it {(0,2) heterotic gauge couplings and their M theory
  origin}},  {\em Nucl.Phys.} {\bf B541} (1999) 109--144,
  [\href{http://xxx.lanl.gov/abs/hep-th/9807124}{{\tt hep-th/9807124}}].

\bibitem{Antoniadis}
I.~Antoniadis, E.~Gava, and K.~Narain, {\it {Moduli corrections to gauge and
  gravitational couplings in four-dimensional superstrings}},  {\em Nucl.Phys.}
  {\bf B383} (1992) 93--109,
  [\href{http://xxx.lanl.gov/abs/hep-th/9204030}{{\tt hep-th/9204030}}].

\bibitem{Vafa}
S.~Cecotti, P.~Fendley, K.~A. Intriligator, and C.~Vafa, {\it {A New
  supersymmetric index}},  {\em Nucl.Phys.} {\bf B386} (1992) 405--452,
  [\href{http://xxx.lanl.gov/abs/hep-th/9204102}{{\tt hep-th/9204102}}].

\bibitem{Moore}
M.~Henningson and G.~W. Moore, {\it {Threshold corrections in K3 x T2 heterotic
  string compactifications}},  {\em Nucl.Phys.} {\bf B482} (1996) 187--212,
  [\href{http://xxx.lanl.gov/abs/hep-th/9608145}{{\tt hep-th/9608145}}].

\bibitem{Gaberdiel}
M.~R. Gaberdiel, S.~Hohenegger, and R.~Volpato, {\it {Symmetries of K3 sigma
  models}},  {\em Commun.Num.Theor.Phys.} {\bf 6} (2012) 1--50,
  [\href{http://xxx.lanl.gov/abs/1106.4315}{{\tt arXiv:1106.4315}}].

\bibitem{Harrison:2013bya}
S.~Harrison, S.~Kachru, and N.~M. Paquette, {\it {Twining Genera of (0,4)
  Supersymmetric Sigma Models on K3}},
  \href{http://xxx.lanl.gov/abs/1309.0510}{{\tt arXiv:1309.0510}}.

\bibitem{04paper}
{\it {Twining Genera and Moonshine of Heterotic Orbifold Compactifications on
  K3}},  {\em to appear}.

\bibitem{CY4paper}
{\it {Does the moon shine on Calabi-Yau fourfolds?}},  {\em to appear}.

\end{thebibliography}\endgroup

\end{document}